\documentclass[floatfix,showpacs,amsmath,amsfonts,amssymb,aps,twocolumn,prb,superscriptaddress, citeautoscript]{revtex4-1}

\usepackage[table,dvipsnames]{xcolor}
\usepackage{graphicx}
\usepackage{dcolumn}
\usepackage{amsthm}
\usepackage{bm}
\usepackage{siunitx}
\usepackage{nicefrac}
\usepackage{mathtools}
\usepackage{float}
\usepackage{braket}
\usepackage{makecell} 
\usepackage{multirow}
\usepackage{wasysym}
\usepackage{titletoc}
\usepackage{multirow}
\usepackage{mwe}
\usepackage{diagbox}

\definecolor{darkred}{rgb}{0.6,0.,0.}
\definecolor{darkgreen}{rgb}{0.,0.5,0.}
\definecolor{darkblue}{rgb}{0.,0.,0.6}
\usepackage[ breaklinks, colorlinks=true, linkcolor=darkred, citecolor=darkgreen, urlcolor=darkblue]{hyperref}

\newcommand{\includegraphicsgood}{\includegraphics}

\usepackage[inactive,tightpage]{preview}
\PreviewMacro[*[[!]{\input{}}
\PreviewMacro[*[[!]{\includegraphicsgood}

\begin{document}

\title{Self-similarity of spectral response functions for fractional quantum Hall states}
\author{Bartholomew Andrews}
\affiliation{Department of Physics, University of Zurich, Winterthurerstrasse 190, 8057 Zurich, Switzerland}
\affiliation{Department of Physics and Astronomy, University of California at Los Angeles, 475 Portola Plaza, Los Angeles, California 90095, USA}
\affiliation{Department of Physics, University of California at Berkeley, 100 South Dr, Berkeley, California 94720, USA}
\author{Gunnar M{\"o}ller}
\affiliation{School of Physical Sciences, University of Kent,  Park Wood Rd, Canterbury CT2 7NZ, UK}
\date{\today}

\begin{abstract}
Spectral response functions are central quantities in the analysis of quantum many-body states, since they describe the response of many-body systems to external perturbations and hence directly correspond to observables in experiments. In this paper, we evaluate a momentum-averaged dynamical density structure factor for the fermionic $\nu=1/3$ fractional quantum Hall state on a torus, using the continued fraction method to compute the dynamical correlation function. We establish the scaling behavior of the screened Coulomb structure factor with respect to interaction range, and expose an inherent self-similarity of structure factors in the frequency domain. These results highlight the statistical properties of spectral response functions for fractional quantum Hall states and show how they can be efficiently approximated in numerical models.
\end{abstract}

\maketitle

\section{Introduction}

One of the key observables yielding insights into interacting quantum systems is the dynamical structure factor $S (\mathbf{q}, \omega)$, which captures the complete momentum- and energy-resolved spectrum of particle excitations. Apart from its central role in the dynamics of quantum many-body systems, the structure factor has a number of appealing properties that stimulate a broad range of research. In particular, we focus on its application in fractional quantum Hall (FQH) systems,  which have been known for several decades to host a rich spectrum of collective modes~\cite{Girvin:1986bu, Lopez93, Haussmann96, Wojs97}, and have been extended to both lattice models~\cite{Repellin14, Wang19} and effective field theories~\cite{Golkar16, Liou19}. Since the structure factor is directly related to the correlation function, it can be computed in a variety of ways, such as via Feynman diagram resummation~\cite{Haussmann96} or continued fractions~\cite{Koch11}. Moreover, the structure factor can be directly probed in two-dimensional electron gases, e.g.~via surface acoustic waves~\cite{Kukushkin09, Wei18}, and analyzed using Raman scattering to reveal additional spin properties~\cite{Nguyen21}. Despite its rich structure and experimental applicability, however, numerical studies that systematically investigate the spectral response of FQH states have only recently gained traction~\cite{Liou19, Kumar22, Repellin14, Nguyen22, Balram22, Wang22, Kumar23}.

In this paper, we study a type of dynamical density structure factor\footnote{We note that the dynamical density structure factor is also known as the density-density response function~\cite{Saarela87} or spectral function of the density operator~\cite{Liou19} in related works.} for the $\nu=1/3$ fermionic Laughlin state on a torus, using the continued fraction method to compute the dynamical correlation function. In particular, two aspects of the structure factor are investigated: (i)~the effect of interaction range, and (ii)~self-similarity. We start by tuning between the $V_1$ and screened Coulomb interactions to reveal the scaling behavior of the structure factor with respect to interaction range. Then, motivated by the fractality of continued fraction Green's functions~\cite{Obata99, Obata99_2}, we study the self-similarity of structure factors for long-range interactions in the frequency domain. In both cases, we present systematic exact diagonalization computations, which we scale with system size. Our results expose the scaling behavior of the structure factor with respect to interaction range, which reflects the functional form of the interaction. Moreover, we reveal that FQH dynamical structure factors are statistically self-similar fractals in the frequency domain, across several orders of magnitude. Apart from providing a deeper insight into the spectral properties of FQH systems, these results may be exploited to compute response functions more efficiently.

The outline of the paper is as follows. In Sec.~\ref{sec:model} we define our FQH system, and in Sec.~\ref{sec:method} we describe the method for computing and analyzing the structure factors. Subsequently, in Sec.~\ref{sec:results} we present our exact diagonalization results. In Sec.~\ref{sec:tune}, we tune the structure factors between the $V_1$ and Coulomb interactions and study the effect of screening. In Sec.~\ref{sec:self}, we examine the self-similarity of the Coulomb structure factor as the frequency domain is rescaled. Finally, in Sec.~\ref{sec:discussion} we discuss the implications with respect to future numerical investigations.

\section{Model}
\label{sec:model}

We consider a two-dimensional system of $N_f$ spin-polarized fermions of mass $m_f$ and charge $q_f$ in a perpendicular magnetic field $B$ on the $xy$-plane with periodic boundary conditions. Building on earlier work~\cite{Keski93, Li93, He94, Wojs97}, the torus geometry has recently experienced a revival of interest~\cite{Pakrouski17, Wang19, Fremling15, Fremling18, Repellin14, Repellin15, Pu17, Pu21}, which motivates our choice. We consider the Landau gauge such that the momentum $k_y$ is a good quantum number. The energy spectrum of this FQH set-up is split into Landau levels, the lowest of which we fill up to a filling factor $\nu = N_f/N_\Phi$, where $N_\Phi$ is the number of flux quanta in the system. Moreover, we focus on the regime where the interaction is weak compared to the Landau level spacing (given by the cyclotron frequency $\omega_\text{c}=q_f B/m_f$). Hence, to a good approximation, we may project the interaction Hamiltonian to the lowest Landau level (LLL), such that
\begin{equation}
\label{eq:ham}
H = H_\text{kin} + \sum_{i<j}^{N_f} P_{\text{LLL}} V(|\mathbf{r}_i - \mathbf{r}_j|) P_{\text{LLL}}, 
\end{equation}    
where $H_\text{kin}$ is the constant kinetic part of the Hamiltonian, $P_{\text{LLL}}$ is the LLL projection operator, $V$ is the interaction potential, and $\mathbf{r}_i$ is the displacement of particle $i$. The relevant length scale in the problem is the magnetic length $l_B=1/\sqrt{q_f B}$. 

In this paper, we consider the Coulomb $V^\text{C}(r)\sim r^{-1}$ and Yukawa $V^\text{Y}_{\lambda}(r)\sim r^{-1}e^{-\lambda r}$ interactions explicitly by diagonalizing the Hamiltonian directly in Fourier space, where $\lambda$ is the Yukawa mass. However, we note that it is not always necessary or desirable to directly account for a long-range interaction in this way. Haldane showed that for systems with a translation and rotation invariant two-body interaction, the interaction Hamiltonian may be written as
\begin{equation}
\label{eq:pp_ham}
H_\text{int}=\sum_{i<j}\sum_L V_L P_{ij}^L,
\end{equation}     
where $V_L$ are the Haldane pseudopotentials, $L$ is the relative angular momentum quantum number between particles $i$ and $j$, and $P_{ij}^L$ is the corresponding projection operator. This simplifies a certain class of long-range interactions into a simple sum of projectors, which has found a diverse set of applications from accelerating early numerical computations on the sphere, through to modeling realistic semiconductor heterojunctions~\cite{Pakrouski17}. Consequently, we complement our analysis by using a Haldane pseudopotential formalism in Sec.~SI of the Supplementary Material. 

Throughout our study, we focus on the primary Laughlin state defined at the filling factor $\nu=1/3$. Laughlin famously proposed a wavefunction ansatz for the ground state of a FQH system with particles interacting via the Coulomb potential in a $1/m$-filled LLL, where $m$ is an odd integer. Although the Laughlin ansatz is a successful description of the problem, since it is in the correct universality class, it is not the exact ground state for the Coulomb interaction. Rather, it was later shown to be the unique, highest-density, zero-energy state for the $V_1$ Haldane pseudopotential. In this paper, we investigate the $\nu=1/3$ state in both limits. When we discuss the ``Laughlin state", we refer to the general ground-state solution to a FQH system with a $1/m$-filled LLL and not the Laughlin ansatz wavefunction in particular.  

\section{Method}
\label{sec:method}

In this section, we outline our numerical method. In Sec.~\ref{sec:sf}, we introduce the continued fraction algorithm for computing dynamical structure factors and in Sec.~\ref{sec:ssp}, we define fractals and self-similar distributions.

\subsection{Structure factors}
\label{sec:sf}

In order to efficiently find the eigenspectrum of the many-body Hamiltonian in Eq.~\eqref{eq:ham}, we employ the Lanczos algorithm~\cite{Koch11}. This method uses an orthogonal Krylov basis, in which the original Hamiltonian $H$ is transcribed to a tridiagonal form $\check{H}$, to compute the eigenbasis:
\begin{align}
H\ket{\Psi_i}&=E_i\ket{\Psi_i}\text{  with  }i=0,\dots,N-1, \\
\check{H}\ket{\check{\Psi}_j}&=\check{E}_j\ket{\check{\Psi}_j}\text{  with  }j=0,\dots, M-1,
\end{align}
where the check marks denote the Krylov representation, $N$ is the dimension of the original Hamiltonian $H$, and $M\leq N$ is the dimension of the Lanczos Hamiltonian $\check{H}$. Tridiagonlization in the Krylov space is rapid, since many degrees of freedom are simultaneously used in the optimization, and memory efficient, since only two vectors of length $N$ need to be stored.\footnote{An additional third vector may be stored to restart the algorithm from a specific point.} Moreover, there is typically good agreement between extremal eigenvalues in the Krylov representation $\check{E}_j$ and those in the original system $E_i$, even for $M \ll N$~\cite{Bai00, Koch11, Dargel12}. Further details of the method are presented in Sec.~SII of the Supplementary Material.

The Lanczos algorithm was later extended by Haydock \emph{et al.}~and applied to compute observables in physical systems with a large number of particles~\cite{Haydock72, Haydock75, Heine80, Bullett80, Haydock80, Kelly80}. In particular, Haydock showed that the resolvent of the Hamiltonian can be efficiently computed using a continued fraction expansion, which is useful for calculating local quantities, such as the single-particle density matrix and the density of states. Crucially, when the original Hamiltonian $H$ is written as a tridiagonal Hamiltonian $\check{H}$ in the Krylov basis, the problem is effectively reduced to a chain of length $M$, which expedites the computation. The algorithm is consequently a widely-used approach in large-scale exact diagonalization computations for quantum many-body systems and has been optimized to diagonalize sparse matrices as large as $\text{dim}(H)\sim10^9$~\cite{Wietek18, Andrews18, AndrewsThesis}.

For our system, we work in momentum space and consider the zero-temperature dynamical correlation function for the operator $O_\mathbf{q}$ in the Lehmann representation, which using the Krylov basis may be approximated as
\begin{align}
\check{G}_{O}(\mathbf{q}, z) &= \sum_{j=0}^{M-1} \frac{|\braket{\check{\Psi}_j|O_\mathbf{q}|\Psi_0}|^2}{E_0 + z-\check{E}_j}\\
&=\braket{\Psi_0|O_\mathbf{q}^\dagger\frac{1}{E_0 + z -\check{H}}O_\mathbf{q}|\Psi_0},
\end{align}    
where $\mathbf{q}\equiv(q_x,q_y)$ are the Fourier components of the $O$ operator, $z\equiv\omega+\mathrm{i}\epsilon$, $\omega$ is the frequency, and $\epsilon$ is a small parameter used to avoid poles in the expansion. From this formula, it is straightforward to show that for the symmetric tridiagonal Hamiltonian $\check{H}$, with ($b_j$)$a_j$ along the (sub)diagonal, the correlation function may be written as a continued fraction
\begin{equation}
\check{G}_{O}(\mathbf{q}, z)=\cfrac{\braket{\Psi_0|O_\mathbf{q}^\dagger O_\mathbf{q}|\Psi_0}}{E_0 + z - a_0 -\cfrac{b_1^2}{E_0 + z-a_1-\cfrac{b_2^2}{\dots}}},
\end{equation}
which terminates at $-b_{M-1}^2 / (z-a_{M-1})$. This form of the correlation function converges rapidly to machine precision~\cite{Dargel12}. 

Specifically, we focus on the density-density correlation functions arising from the density operator
\begin{equation}
\rho_\mathbf{q}\equiv \int \,\mathrm{d}\mathbf{r}\, e^{\mathrm{i}\mathbf{q}\cdot\mathbf{r}} c^\dagger(\mathbf{r}) c(\mathbf{r}),
\end{equation}
where $\mathbf{r}\equiv(x,y)$ is the position operator conjugate to $\mathbf{q}$. Given our choice of Landau gauge with definite momentum $k_y$, we are particularly interested in resolving the $q_y$ Fourier components  of the density operator. We therefore choose to integrate out the $q_x$ modes on the torus to avoid an additional free variable and consider the $q_x$-momentum-averaged density operator, setting
\begin{equation}
O_{q_y}\equiv \bar \rho_{q_y}=\sum_{m=0}^{N_\Phi-1} \rho_{q_x=\frac{2\pi m}{L_x}, q_y},
\end{equation}
where $L_x\times L_y$ are the system dimensions. We have separately verified, by evaluating at specific $q_x$ values, that the density operator is only weakly dependent on $q_x$. The full derivation of the momentum-averaged density operator is presented in Sec.~SIII of the Supplementary Material. Finally, we may use this operator to compute the corresponding dynamical density structure factor
\begin{equation}
\label{eq:spec_func}
\check{I}_{\bar\rho}(q_y,\omega) = -\frac{1}{\pi} \lim_{\epsilon\to 0} \text{Im}\, \check{G}_{\bar\rho}(q_y, \omega + \mathrm{i} \epsilon),
\end{equation}
which we often refer to simply as the ``structure factor". The crucial property of the continued fraction expansion is that the structure factor in the Krylov representation $\check{I}$ accurately reproduces the moments of the structure factor in the Hilbert representation $I$, and so we now drop the check marks~\cite{Koch11}.

\subsection{Fractals and self-similarity}
\label{sec:ssp}

In this work, we investigate the self-similarity of the structure factors $I_{\bar{\rho}}(q_y, \omega)$. Fractals and self-similarity appear in many contexts in condensed matter physics, such as the Hofstadter spectrum of energy levels for electrons hopping in a periodic potential~\cite{Hofstadter76}, the Haldane hierarchy of stable FQH filling fractions~\cite{Haldane83}, and the statistical analysis of time series~\cite{Weron05}. Moreover, they have several important characteristics that can often be leveraged in theory and simulations.

A fractal is defined as an object with a fractal dimension $D$ that is greater than its topological dimension $d$~\cite{HastingsBook}. The fractal dimension may be computed in a variety of ways, however is traditionally defined via $n\equiv s^{-D}$, where $n$ is the number of units in the whole object and $s$ is the scale factor. One of the distinctive properties of fractals is their scale-invariance, also known as self-similarity, where subregions of a structure are identical to the whole. However, we note that not all self-similar objects are fractals. For example, a square is a self-similar object with $D=d=2$, whereas a Koch curve is a fractal with $D=\log(4)/\log(3)>d$~\cite{Falconer14}.

As for the fractional dimension above, self-similarity may also be defined differently depending on the context. Exact self-similarity holds on all scales, and in this case the various definitions of the fractal dimension coincide. However, quasi or multi-fractal self-similarity is more common, with lower and upper bounds on where this behavior applies. In functional analysis, self-similarity occurs when a subsection of a function statistically resembles the entire function. Specifically, for a function of one variable $I(\omega)$, this occurs when
\begin{equation}
\label{eq:ss}
I(\omega)\equiv s^\kappa I\left(\frac{\omega}{s}\right),
\end{equation}             
where $s$ is a scale factor and $\kappa$ is the self-similarity parameter~\cite{Embrechts00}. In Eq.~\eqref{eq:ss}, ``$\equiv$" implies that the distributions on both sides of the equation are statistically identical. However, in practice, this is approximated by examining the first and second moments~\cite{Embrechts00, Peng00, Mandelbrot10}. 

In contrast to geometric analysis, a general function of one variable is in a two-dimensional space where each axis represents different physical quantities. Consequently, two magnification factors are required to quantify self-similarity, such that
\begin{equation}
\label{eq:alpha}
\kappa \equiv \frac{\log M_y}{\log M_x},
\end{equation}
where $M_i$ is the magnification factor of the $i$-axis. Note that this takes an analogous form to the definition of fractal dimension discussed above, albeit with a different interpretation. The fractional dimension of a function is often difficult to quantify. However, since the demonstration of self-similarity for any non-trivial curve indicates detail across many orders of magnitude, which precludes an integer dimension, this is taken as evidence to show that a curve is a fractal with respect to the axes on which the magnification occurs.

\section{Results}
\label{sec:results}

In this section, we present our exact diagonalization results~\cite{sr_data}. In Sec.~\ref{sec:tune}, we investigate the scaling of structure factors as we tune from the $V_1$ to the screened Coulomb interaction, and in Sec.~\ref{sec:self}, we expose a statistical self-similarity of structure factors in the frequency domain.

\subsection{Tuning the interaction range}
\label{sec:tune}

We compute the momentum-averaged dynamical density structure factor $I_{\bar{\rho}}$ for the $\nu=1/3$ Laughlin state stabilized by a linear superposition of the $V_1$ Haldane pseudopotential~\cite{Haldane83} and an explicit $V^\text{Y}_\lambda(q)$ Yukawa interaction. In this system, the interaction Hamiltonian is given as $H_\text{int}=(1-\alpha)H_{V_1}+\alpha H_{V^\text{Y}_\lambda}$, where $\alpha\in[0, 1]$. The tuning parameter $\alpha$ allows us to interpolate between two common ground-state solutions in the same universality class, and the Yukawa mass $\lambda$ enables us to vary the interaction range and recover the Coulomb limit.

\begin{figure}
	\includegraphics[width=\linewidth]{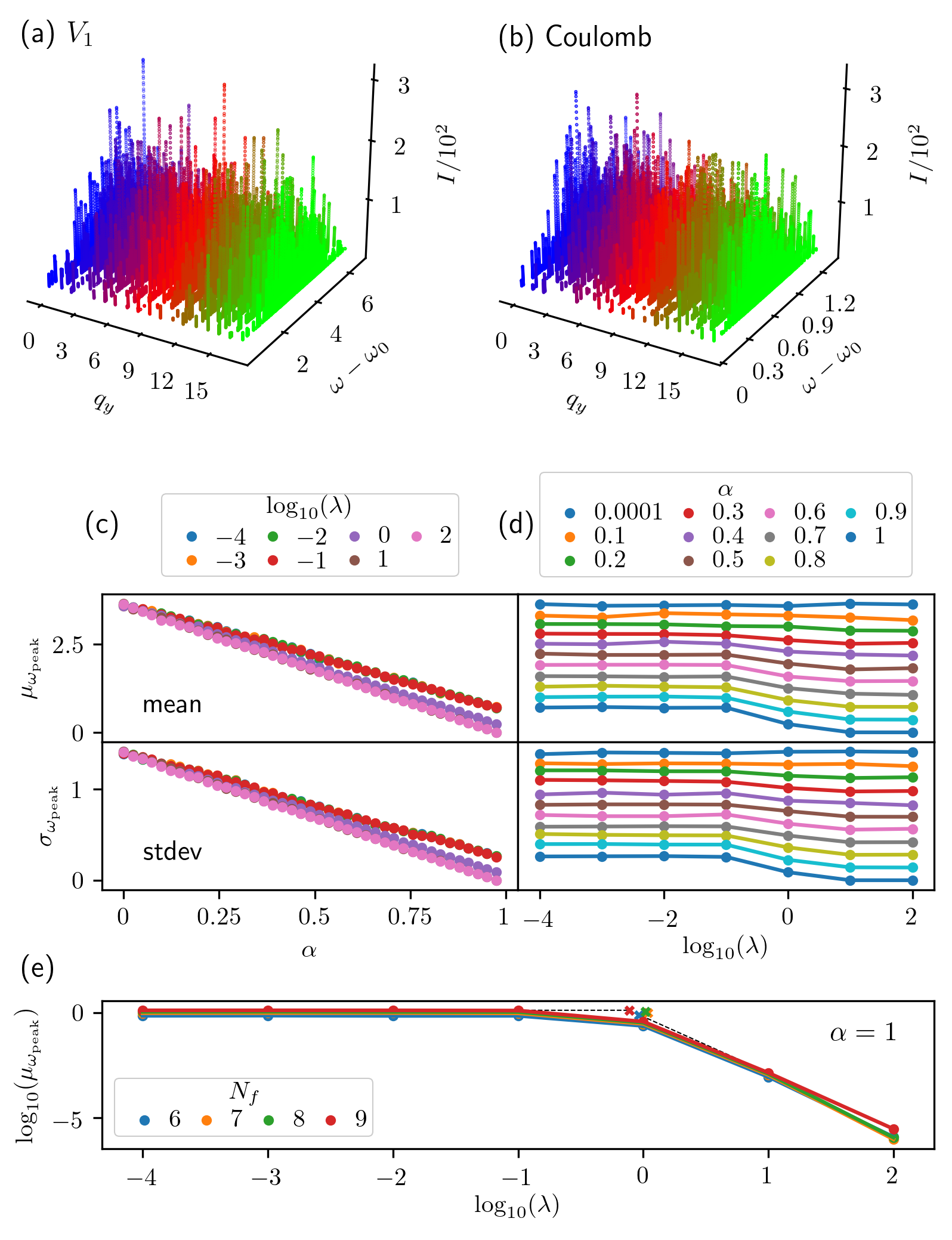}
	\caption{\label{fig:sr_lty}\textbf{Tuning the structure factor with respect to interaction range.} Structure factors $I_{\bar{\rho}}(q_y,\omega)$ as a function of angular frequency offset by the ground state energy $\omega-\omega_0$, for the $\nu=1/3$ FQH state on a torus, with $N_f=6$ particles and $N_\Phi=18$ flux quanta, stabilized by (a)~the $V_1=1$ pseudopotential, (b)~the exact Coulomb interaction $V^\text{C}$, and \mbox{(c--e)}~$H_\text{int}=(1-\alpha)H_{V_1}+\alpha H_{V^\text{Y}_\lambda}$, in the LLL. In (a,b), the spectra are resolved with respect to their $q_y$ momentum sector, whereas in (c--e) $q_y=0$. The mean, $\mu$, and standard deviation, $\sigma$, of offset angular frequencies coinciding with spectral peaks, $\omega_\mathrm{peak}$, are shown as a function of (c)~$\alpha$ and (d)~$\lambda$. (e)~Finite-size scaling of the $\alpha=1$ curve from (d). The transition points between the two regimes are marked with crosses. The computations were performed with a resolution of (a,b)~$\Delta\omega=10^{-5}$, $\Delta I =10^{-5}$, $\epsilon=10^{-4}$, and (c--e)~$\Delta\omega=10^{-7}$, $\Delta I =10^{-4}$, $\epsilon=10^{-6}$.}
\end{figure}

In Figs.~\ref{fig:sr_lty}(a,b), we start by computing the structure factors corresponding to the two most common approaches for stabilizing the $\nu=1/3$ Laughlin state, via the $V_1$ and Coulomb interactions. We present our initial results for all $q_y$ momentum sectors. Since the structure factor defined in Eq.~\eqref{eq:spec_func} conserves particle number, these plots show the coupling of the ground state to gapped excitations and consist of a spectrum of peaks at finite frequency. As expected: the structure factor corresponding to the $V_1$ interaction yields a broader spread of frequencies, due to the normalization of the $V_1=1$ pseudopotential~\cite{Nguyen22}; the relative peak amplitudes are consistent in the two cases, owing to the dominant $V_1$ component of the Coulomb interaction~\cite{Wang22, Kumar22}; and the shape of both distributions is unimodal, according with the theory for Laughlin states~\footnote{In the literature, these collective peaks are typically $q_x$-momentum-resolved and may sometimes be referred to simply as peaks~\cite{Nguyen22, Balram22}, or collective modes~\cite{Wang22, Kumar22} in certain contexts.}. Up to slight variations in the number and heights of the peaks, the overall shape of the envelope holds for all runs and for all $q_y$, and there is a close resemblance between the structure factors of these two FQH states.

Motivated by the effect that interaction range has on the form of the structure factors, in Figs.~\ref{fig:sr_lty}(c,d) we tune $I_{\bar{\rho}}(0,\omega)$ from Figs.~\ref{fig:sr_lty}(a,b), with respect to $\alpha$ and $\lambda$, at the increased resolution of $\Delta\omega=10^{-7}$ and $\epsilon=10^{-6}$.~\footnote{Although the evolution is shown only for $q_y=0$, this analysis holds for all momentum sectors.} We note that decreasing $\epsilon$ has the auxiliary effect of proportionally increasing the peak amplitudes and decreasing the peak widths. We present the evolution of the first two moments of the distribution: the mean (top panels) and standard deviation (bottom panels). In this case, we consider the offset angular frequencies coinciding with spectral peaks, $\omega_\mathrm{peak}$, and use their mean, $\mu_{\omega_\mathrm{peak}}$, and standard deviation, $\sigma_{\omega_\mathrm{peak}}$, as quantifiers of center and spread, respectively.~\footnote{In Fig.~\ref{fig:sr_lty}, we chose to study the self-similarity of $\{\omega_\mathrm{peak}\}$ with respect to the $\alpha$ and $\lambda$ axes. However, analogous relations also hold for $\{I_\mathrm{peak}\}$.} 

From the constant gradient of the first two moments in Fig.~\ref{fig:sr_lty}(c), we can see that the structure factor scales linearly with the tuning parameter $\alpha$. This is expected, since we are effectively multiplying the Yukawa interaction by a scale factor modulo a correction from the $V_1$ term. Subsequently, in Fig.~\ref{fig:sr_lty}(d), we plot the scaling of the structure factor on different axes, to clearly show the influence of $\lambda$. As $\alpha\to 0$, the structure factor does not depend on $\lambda$, since there is a vanishing component of the Yukawa interaction in the Hamiltonian. Similarly, the influence of $\lambda$ increases linearly with $\alpha$. Most notably, however, we observe two non-trivial scaling regimes for the structure factor as $\alpha\to 1$. For $\log\lambda \lesssim -1$, the structure factor is approximately independent of $\lambda$, whereas for $\log\lambda > -1$, the center and spread exponentially diminish to zero. This behavior reflects the exponential suppression of the Yukawa interaction potential at large Yukawa mass, which correspondingly restricts the domain of the response functions.

To investigate this transition in detail, in Fig.~\ref{fig:sr_lty}(e) we illustrate the finite-size scaling of the $\alpha=1$ curve from the top panel of Fig.~\ref{fig:sr_lty}(d) on a log-log plot. Here, we see explicitly that the continuous connection between $\alpha$ and $\lambda$ translates to a non-trivial scaling with respect to $\lambda$, with two regimes. Connecting lines of best fit from these two regions, yields a transition point at $\log\lambda\approx0$. Using Eq.~\eqref{eq:alpha}, with $M_x$ corresponding to $\lambda$ and $M_y$ corresponding to $\mu_{\omega_\text{peak}}$, we obtain the self-similarity parameters $\kappa_\mu= 0.00116$ and $-2.67$ for $\log\lambda\lesssim 0$ and $\log\lambda>0$, respectively. This reflects the asymptotic scaling of the Yukawa interaction potential in the small and large $\lambda$ limits. Note that we used a linear scale for $\alpha$ in Fig.~\ref{fig:sr_lty}(c), since this corresponds to linearly interpolating between two Hamiltonians, whereas we use a logarithmic scale for $\lambda$, to analyze a wide scope of interaction ranges~\footnote{The interaction matrix elements are too small for $\lambda\geq10^3$ to reliably stabilize the fermionic Laughlin state.}.    

In this section, we have established the scaling of structure factors with respect to $\alpha$ and $\lambda$ in the framework of statistical self-similarity, and showed that it reflects the functional form of the interaction. Moreover, the scaling is not \emph{exactly} self-similar, since the combined effect of peak fluctuations due to microscopic details of the Hamiltonian, and numerical noise due to sample aliasing, yields $\sim1\%$ fluctuations in peak number and amplitude. 

\subsection{Rescaling the frequency domain}
\label{sec:self}


Previously, we demonstrated that the structure factor scales trivially with respect to a linear interpolation between the $V_1$ and screened Coulomb interactions, and non-trivially with respect to interaction range, reflecting the functional form of the interaction. In both cases, this scaling is the result of tuning parameters; namely, $\alpha$ and $\lambda$. In this section, we investigate a form of self-similarity with respect to the frequency domain, which cannot be explicitly linked to a tuning parameter. 

\begin{figure}
	\includegraphics[width=\linewidth]{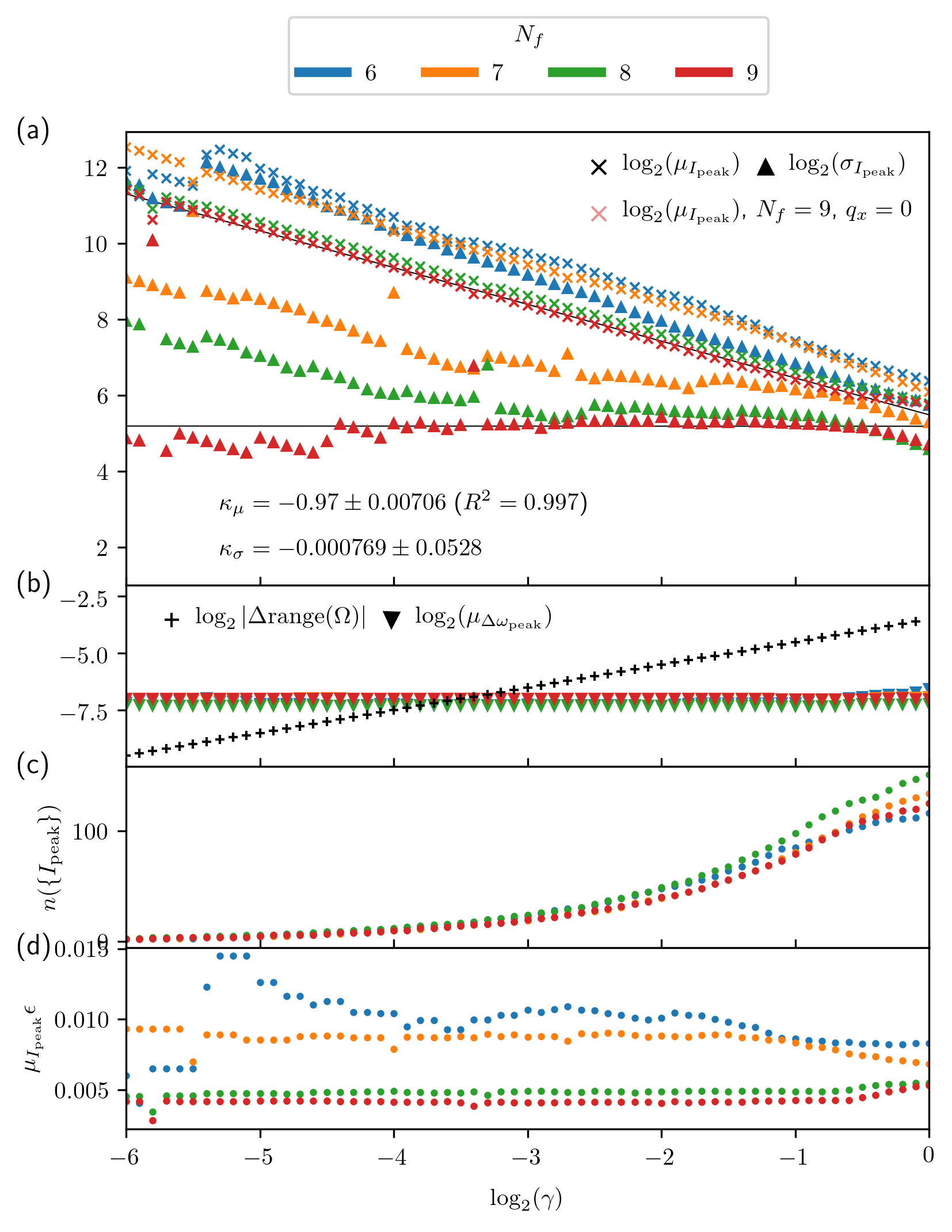}
	\caption{\label{fig:coulomb_scaling}\textbf{Rescaling the structure factor in the frequency domain.} Structure factor $I_{\bar{\rho}}(0, \omega)$ for the $\nu=1/3$ FQH state on a torus, with $N_f=6,7,8,9$ particles, stabilized by the exact Coulomb interaction $V^\text{C}$, in the LLL. For comparison, we overlay the $N_f=9$ data at $q_x=0$ with half opacity. (a)~Mean, $\mu$, and standard deviation, $\sigma$, of the peaks of the structure factor, $I_\mathrm{peak}$, as we symmetrically scale the $\Omega_0$ domain about its midpoint $\omega_\text{mid}=(\omega_\text{min}+\omega_\text{max})/2$, by a scale factor $\gamma\equiv\text{range}(\Omega)/\text{range}(\Omega_0)$. The initial frequency domain, $\Omega_0$, is chosen to span the entire structure factor. For each iteration, we correspondingly scale $\epsilon$, to reduce the widths of the peaks, and $\Delta\omega$ to increase our numerical resolution. The lines of best fit for the $N_f=9$ data in the linear regions are drawn in black and the self-similarity parameters are given by the gradients of the slopes. (b)~The magnitude of the $\Omega$ domain reduction between successive steps, $\Delta\text{range}(\Omega_i)\equiv\text{range}(\Omega_{i})-\text{range}(\Omega_{i-1})$, where $i$ is the frequency domain index, and the average separation between $\omega_\mathrm{peak}$ values, $\mu_{\Delta \omega_\text{peak}}$. (c)~The number of peaks, $n(\{I_\mathrm{peak}\})$, and (d)~the average peak magnitude, $\mu_{I_\mathrm{peak}}$, as we scale $\Omega$. The first computation with $\Omega=\Omega_0$ was performed with a resolution of $\Delta\omega=10^{-5}$, $\Delta I =10^{-5}$, and $\epsilon=10^{-4}$.}
\end{figure}

In Fig.~\ref{fig:coulomb_scaling}, we investigate the distribution of the peak magnitudes, $I_\text{peak}$, in the Coulomb structure factor from Fig.~\ref{fig:sr_lty}(b), as we scale the frequency domain. For clarity, we denote angular frequencies with a lower-case $\omega$ and a set of angular frequencies with an upper-case $\Omega\equiv\{\omega\}$. We consider an initial frequency domain $\Omega_0$ with $\text{range}(\Omega_0)=\omega_\text{max}-\omega_\text{min}$, which we scale symmetrically about its midpoint $\omega_\text{mid}=(\omega_\text{min}+\omega_\text{max})/2$, by a scale factor $\gamma \equiv \text{range}(\Omega)/\text{range}(\Omega_0)$. We choose $\Omega_0$ to span the entire structure factor, although we note that the precise choice is arbitrary. In order to keep the scaling numerically consistent, we correspondingly scale the frequency resolution, $\Delta\omega$, and the $\epsilon$ value in our simulations.
%
In Fig.~\ref{fig:coulomb_scaling}(a), we plot the first two moments of the distribution against the domain scale factor to compute the self-similarity parameters. We additionally scale this with system size up to $N_f=9$ particles. We find that there is a contiguous linear region, which grows with system size, and has a correlation coefficient of $R^2>0.99$~\footnote{The $R^2$ value is defined as the square of Pearson’s correlation coefficient for the lines of best fit. Note that $R^2$ cannot be computed when the gradient $\kappa_{\sigma}\approx0$.}, which indicates a statistical self-similarity of the Coulomb structure factor with respect to frequency domain rescaling. This behavior also holds for the $q_x$-momentum-resolved dynamical density structure factor, as demonstrated for $N_f=9$, $q_x=0$. The self-similarity parameters for the mean and standard deviation are $\kappa_\mu=-0.97\pm0.00706$ and $\kappa_\sigma=-0.000769\pm0.0528$. As mentioned in Sec.~\ref{sec:tune}, since a reduction of $\epsilon$ increases the heights of the peaks, it is consistent that the self-similarity parameter $\kappa_{\mu}\approx -1$. Finite-size scaling shows that the self-similarity parameter for the standard deviation holds deeper into the domain rescaling procedure with increasing system size, and maintains a constant value across the entire procedure for $N_f=9$, up to sampling effects at small $\Omega$. 
%
In Fig.~\ref{fig:coulomb_scaling}(b), we compare the magnitude of the frequency interval reduction between successive steps, $\Delta\text{range}(\Omega_i)\equiv\text{range}(\Omega_{i})-\text{range}(\Omega_{i-1})$, where $i$ is the frequency domain index, with the average spacing between the peaks, $\mu_{\Delta\omega_\text{peak}}$. Since the frequency domain scaling requires a finite section of structure factor peaks to be truncated on each iteration, this allows us to verify the continuity of the procedure. We observe that the average spacing between the peaks is initially smaller than the size of the frequency interval being removed, up to $\gamma\approx 2^{-3.5}$. For smaller values of $\gamma$, breakdowns in the rescaling continuity become more likely, as observed for the $\sigma_{I_\text{peak}}$, $N_f=9$ data in Fig.~\ref{fig:coulomb_scaling}(a). However, the average spacing between the peaks remains constant, independent of particle number, which shows that this is not the cause of numerical breakdown for small system sizes.
%
In Fig.~\ref{fig:coulomb_scaling}(c), we examine the number of peaks in the structure factor, $n(\{I_\text{peak}\})$, with frequency domain rescaling. Since the moments of a distribution are sensitive to the sample size, this is another validity test for the self-similar scaling. We find that the number of peaks in the structure factor decreases exponentially with frequency domain reduction, which holds independently of system size, up to the influence of initial conditions at $\Omega\approx\Omega_0$. Nevertheless, for $\gamma\lesssim 2^{-5.5}$, we find that for smaller particle numbers, there are slightly fewer peaks in the spectrum, which may be a contributing cause of the numerical breakdown for small system sizes, since the number of peaks is already extremely low.
%
Finally, in Fig.~\ref{fig:coulomb_scaling}(d), we analyze the average magnitude of spectral peaks, $\mu_{I_\text{peak}}$, as we shrink the frequency domain. Since the height of the peaks increases linearly with decreasing $\epsilon$, we expect that $\mu_{I_\text{peak}}\epsilon$ is constant in the linear region. This holds approximately for larger system sizes, up to initial conditions at $\Omega\approx\Omega_0$. However, for smaller systems, and particularly $N_f=6$, we see that the mean amplitude of the peaks fluctuates during the procedure, which shows that the influence of peak fluctuations and numerical noise is too great for a reliable scaling.

In general, structure factors for the Yukawa interaction are statistically self-similar with respect to frequency domain rescaling, for all values of screening. This is quantified by the linear scaling of their first two moments, as shown for $\lambda=0$ in Fig.~\ref{fig:coulomb_scaling}. However, since the mean and standard deviation of the structure factors approaches zero as $\lambda\to\infty$, as shown in Fig.~\ref{fig:sr_lty}, this self-similarity is most apparent for long-range interactions.

\section{Discussion and conclusions}
\label{sec:discussion}

In this paper, we studied numerically the momentum-averaged dynamical density structure factors $I_{\bar{\rho}}$ (Eq.~\eqref{eq:spec_func}) for the $\nu=1/3$ Laughlin state on the torus, using the continued fraction method. Our main result is the discovery of a statistical self-similarity of the structure factor in the frequency domain. Specifically, in Sec.~\ref{sec:self}, we showed that the peak distribution has fractal properties across several orders of magnitude. This self-similar nature is realized most precisely and across the largest range of frequency scales in the thermodynamic limit, for systems stabilized by long-range interactions. In addition, in Sec.~\ref{sec:tune}, we established the scaling behavior of the structure factor with respect to interaction range. This dependence is determined by the asymptotic behavior of the Yukawa interaction with respect to the screening parameter $\lambda$.

Physically, the structure factor $I_{\bar{\rho}}(q_y, \omega)$, with $q_y$ fixed, corresponds to the energy-resolved spectrum of particle excitations. The amplitudes and fine structure in the spectrum of peaks are consequently related to the probability distribution of many-particle excitations in the system. In Sec.~\ref{sec:tune}, we showed that an increase in the interaction range proportionally increases the center and spread of particle excitations, which is a direct result of the increased average interaction amplitude, as well as the number of interaction permutations. Furthermore, in Sec.~\ref{sec:self}, we demonstrated the fractality of the structure factors with respect to the energy axis, which reflects the continued chain of possible many-particle interactions with diminishing amplitudes, and is especially prevalent for systems stabilized by long-range interactions. Building on this, we expect that these properties also hold in the dynamical structure factors of other strongly correlated phases of matter with Coulomb-type interactions, such as superconductors~\cite{Dutta13} or transition metal oxides~\cite{Gautreau21}.

Our results highlight the effect of interaction range on, and the self-similarity of, dynamical structure factors $I_{\bar{\rho}}$ for FQH systems. However, statistical self-similarity is a more general property of response functions in condensed matter systems, and beyond. In particular, there have been a wealth of studies on the self-similarity, fractality, and chaos of continued fractions in a mathematical context~\cite{Obata99, Obata99_2}, and this is reflected in a wide class of observables derived from the Green's function. We have explicitly demonstrated the fine structure of spectral response functions in the frequency domain, which stems in part from their continued fraction representation. This leaves scope for further manifestations of self-similarity due to this recurrence relation, as well as potential applications. For example, there is a natural limitation to the achievable energy resolution of structure factors derived from experiments, such as inelastic x-ray scattering and photoemission spectroscopies~\cite{Seidu18, Ruotsalainen21}, which may be numerically enhanced using statistical interpolation. Moreover, dynamical quantum simulators have recently been proposed as a method to compute structure factors~\cite{Baez20, Sun23}, which may be expedited using self-similarity relations. On a more pragmatic level, our results offer a way to efficiently approximate the Coulomb structure factor. Specifically, for systems stabilized with long-range interactions, the structure factor may be readily derived by diagonalizing a short-range Yukawa interaction Hamiltonian in Fourier space. Large $\lambda$ yields a short-range interaction that is efficient to implement, and, provided the simulation resolution is sufficiently high, this result can be smoothly tuned to the long-range Coulomb limit.

To complement these results, in the Supplementary Material we examined the behavior of dynamical structure factors for FQH states that are stabilized by Haldane pseudopotential interactions, which contrasts the effects of tuning the interaction range via the Yukawa mass, and truncating two-body interactions with large relative angular momenta. We showed that Haldane pseudopotentials are not designed to model long-range interactions on the torus at the system sizes currently accessible, however a reasonable approximation may be achieved, provided that the interactions are modulated to be sufficiently short-range relative to the system size. Using the example of the Coulomb structure factor from Fig.~\ref{fig:sr_lty}(b), we found that the optimal approximation was recovered at pseudopotential order $\beta\sim N_\Phi / 2$ with a weakly screened form of the interaction. This demonstrates that, provided sufficient care is taken, Haldane pseudopotentials provide another route to approximate dynamical structure factors for systems stabilized with long-range interactions, at a significantly reduced numerical cost.

There are several ways in which this work could be extended in the future. First, it would be interesting to build on this analysis of ground states in the same universality class at $\nu=1/3$, to other FQH filling factors, and in particular, ground states that do not share a universality class and inherently require a long-range interaction to be stabilized~\cite{Yang19, Andrews21}. Second, it would be useful to analyze the trade-off between interaction range / frequency window and simulation resolution using this approach to find the optimal efficiency benefit for a series of FQH states. Finally, there is current motivation to leverage this method and compute the full density-density response function, to identify collective excitations in FQH systems, and guide the latest experimental techniques, including spin wave spectroscopy in graphene~\cite{Wei18}.

\begin{acknowledgments}
We thank Garry Goldstein for help with the derivation of the momentum-averaged density operator, detailed in Sec.~SIII of the Supplementary Material. We thank Steve Simon, Nigel Cooper, Zlatko Papic, Nicholas Regnault, Kiryl Pakrouski, Titus Neupert, and Jie Wang for useful discussions. Our numerical results were produced using \textsc{DiagHam}. B.A.~acknowledges support from the Swiss National Science Foundation under Grant Nos.~P500PT\_203168 and~PP00P2\_176877, as well as the University of California Laboratory Fees Research Program funded by the UC Office of the President (UCOP), Grant No.~LFR-20-653926. G.M.~acknowledges support from the Royal Society under University Research Fellowship URF\textbackslash R\textbackslash180004.
\end{acknowledgments}

\bibliographystyle{apsrev4-1}
\bibliography{ms}

\end{document}



\onecolumngrid


\begin{center}
	\textbf{\large Supplementary Material: Self-similarity of spectral response functions for fractional quantum Hall states}
\end{center}

\tableofcontents

\setcounter{equation}{0}
\setcounter{figure}{0}
\setcounter{table}{0}
\setcounter{section}{0}
\makeatletter
\renewcommand{\theequation}{S\arabic{equation}}
\renewcommand{\thefigure}{S\arabic{figure}}
\renewcommand{\thetable}{S\arabic{table}}
\renewcommand{\thesection}{S\Roman{section}}
\renewcommand{\bibnumfmt}[1]{[S#1]}
\renewcommand{\citenumfont}[1]{S#1}

\section{Haldane pseudopotential formalism}
\label{sec:hal}

In this section, we discuss the properties of FQH systems with interactions modeled using Haldane pseudopotentials. In Sec.~\ref{sec:pp_der}, we outline the derivation of Haldane pseudopotentials for the Coulomb and Yukawa interactions; in Sec.~\ref{sec:pp_conv}, we benchmark the convergence of two-particle energy spectra; and in Sec.~\ref{sec:trunc}, we analyze the effect on spectral response functions as we tune the interaction from lower- to higher-order pseudopotentials. Although the examples studied can be analyzed directly using exact diagonalization, we use a Haldane pseudopotential approach to contrast the effects of tuning the interaction range via the Yukawa mass, as shown in Sec.~IV.A, and truncating two-body interactions with large relative angular momenta.

\subsection{Derivation of the Haldane pseudopotentials}
\label{sec:pp_der}

The derivation in this section is performed on a plane and in atomic units, where the Coulomb constant $k_e\equiv(4\pi\epsilon)^{-1}=1$, the fermionic charge $q_f=1$, and the magnetic length $l_B\equiv(2\pi N_\Phi)^{-1/2}=1$.

The Haldane pseudopotentials~\cite{Haldane83} in the $n$-th Landau level, $V_m^{(n)}\equiv \braket{n,m|V|n,m}$, may be written in momentum space as
\begin{equation}
	V_m^{(n)}=\int \frac{\mathrm{d}\mathbf{q}}{2\pi} V(\mathbf{q}) \braket{n,m|e^{\mathrm{i}\mathbf{q}\cdot\mathbf{r}}|n,m},
\end{equation}
where $m$ is the pseudopotential index and $\mathbf{r}$, $\mathbf{q}$ are the position and momentum vectors, respectively. 

Using the fact that the FQHE in the $n$-th Landau level with interaction potential $V(\mathbf{q})$ is equivalent to the FQHE in the 0-th Landau level with interaction potential $[L_n(q^2/2)]^2V(\mathbf{q})$, coupled with the result $\braket{m|e^{\mathrm{i}\mathbf{q}\cdot\mathbf{r}}|m}=e^{-q^2}L_m(q^2)$, allows us to write
\begin{equation}
	V_m^{(n)}=\int \frac{\mathrm{d}\mathbf{q}}{2\pi} V(\mathbf{q}) \left[L_n (q^2/2) \right]^2 L_m(q^2) e^{-q^2},
\end{equation}
where $L_n$ is the $n$-th Laguerre polynomial~\cite{MacDonald94, Jain07}.

Finally, converting the momentum integral to polar coordinates yields
\begin{equation}
	\label{eq:haldane}
	V_m^{(n)}=\int_0^{\infty} q\mathrm{d} q V(\mathbf{q}) \left[L_n (q^2/2) \right]^2 L_m(q^2) e^{-q^2}.
\end{equation}

\subsubsection{Coulomb interaction}

We use the explicit representation of a centrally-symmetric interaction $V(\mathbf{r}) = V(r)$ in momentum space,
\begin{equation}
	V(q)=\int \frac{\mathrm{d}\mathbf{r}}{2\pi} V(r) e^{-\mathrm{i}\mathbf{q}\cdot\mathbf{r}}.
\end{equation}
Converting this momentum integral to polar coordinates, coupled with the fact that $e^{-\mathrm{i}\mathbf{q}\cdot\mathbf{r}}=\cos(\mathbf{q}\cdot\mathbf{r})$ since the interaction is a real and even function of $r$, allows us to write
\begin{equation}
	V(q)=\int_0^{\infty} r\mathrm{d}r V(r) \int_0^{2\pi} \frac{\mathrm{d}\theta}{2\pi} \cos(qr\cos \theta).
\end{equation}
Furthermore, since the angular integral is a standard Bessel integral, this expression reduces to
\begin{equation}
\label{eq:VofQ}
	V(q)=\int_0^{\infty} r\mathrm{d}r V(r) J_0(qr),
\end{equation}
where $J_n$ is the $n$-th Bessel function of the first kind. 

Specifically for the Coulomb interaction, $V^\text{C}(r)= r^{-1}$, this yields
\begin{equation}
	V^\text{C}(q)=\int_0^{\infty} r\mathrm{d}r V^\text{C}(r) J_0(qr) =\int_0^{\infty}\mathrm{d}r J_0(qr) = \frac{1}{q}.
\end{equation}
Inserting this result into the expression for the Haldane pseudopotentials, Eq.~\eqref{eq:haldane}, then yields 
\begin{equation}
	V_m^{C,(n)}=\int_0^{\infty} \mathrm{d}q \left[L_n (q^2/2) \right]^2 L_m(q^2) e^{-q^2}.
\end{equation}
Finally, in the LLL ($n=0$), this integral can be evaluated analytically and written in closed form, such that
\begin{equation}
	\label{eq:coulomb_pp}
	V_m^{C,(0)}=\frac{\sqrt{\pi}}{2}\prescript{}{2}{F}_1(1/2,-m;1;1) = \frac{\sqrt{\pi}}{2}\frac{(2m-1)!!}{2^m m!},
\end{equation}
where $\prescript{}{2}{F}_1$ is the Gauss hypergeometric function. The asymptotic scaling as $m\to\infty$ is given as
\begin{equation}
	V_m^{C,(0)}\sim\frac{2^{-5/4+\cos(2m\pi)/4}\pi^{\sin^2(m\pi)/2}}{\sqrt{m}}.
\end{equation}

\subsubsection{Yukawa interaction}

\begin{figure}
	\includegraphics[width=.7\linewidth]{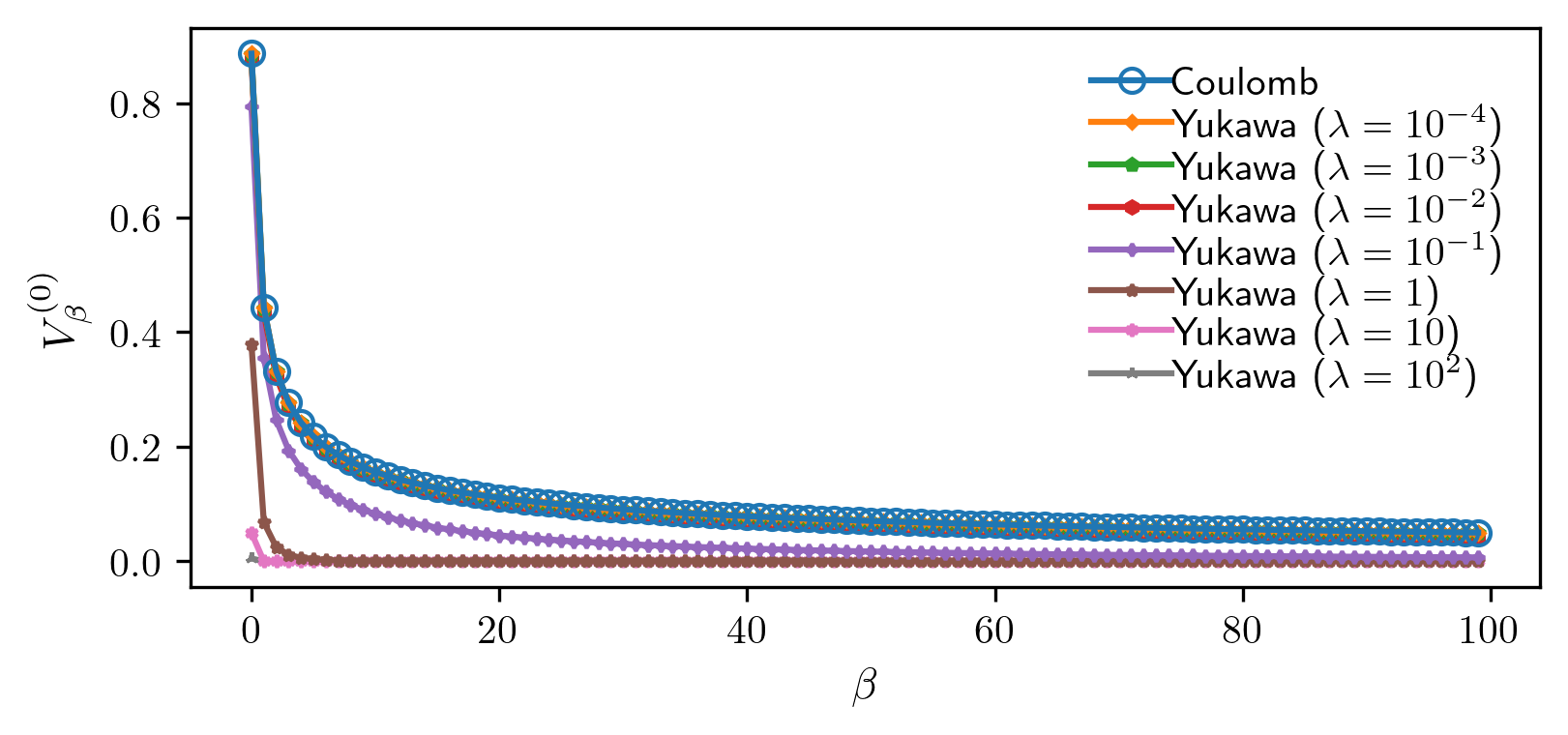}
	\caption{\label{fig:pp}\textbf{Haldane pseudopotentials.} Haldane pseudopotentials on a plane, in the LLL, for the Coulomb~\eqref{eq:coulomb_pp} and Yukawa interactions~\eqref{eq:yukawa_pp}.}
\end{figure}

Similarly, using \eqnref{VofQ}, the Yukawa interaction, $V^\text{Y}_{\lambda}(r)= r^{-1}e^{-\lambda r}$, may be written in momentum space as
\begin{equation}
	V^\text{Y}_\lambda(q)=\int_0^{\infty} r\mathrm{d}r V^\text{Y}_{\lambda}(r) J_0(qr)=\frac{1}{\sqrt{\lambda^2+q^2}},
\end{equation}
where $\lambda$ is the Yukawa scaling constant. Inserting this result into the expression for the Haldane pseudopotentials, Eq.~\eqref{eq:haldane}, yields
\begin{equation}
	V_{\lambda,m}^{Y,(n)}=\int_0^{\infty} \mathrm{d}q \frac{q}{\sqrt{\lambda^2 + q^2}} \left[L_n (q^2/2) \right]^2 L_m(q^2) e^{-q^2}.
\end{equation}
In the LLL, this expression simplifies to
\begin{equation}
	\label{eq:yukawa_pp}
	V_{\lambda,m}^{Y,(0)}=\frac{\lambda}{2\sqrt{\pi}} \Gamma(m+1/2)U(m+1,3/2,\lambda^2),
\end{equation}
where $\Gamma$ is the gamma function and $U$ is the confluent hypergeometric function of the second kind. The asymptotic scaling as $m\to\infty$ is given as
\begin{equation}
	V_{\lambda,m}^{Y,(0)}\sim \frac{\lambda}{\sqrt{2}} e^{-m} m^m U(m+1,3/2,\lambda^2).
\end{equation}
Plots of the Haldane pseudopotentials for both the Coulomb and Yukawa interactions are shown in Fig.~\ref{fig:pp}.

\subsection{Convergence of two-particle energy spectra}
\label{sec:pp_conv}

One of the fundamental properties of Haldane pseudopotentials is their correspondence with the two-particle energy spectrum~\cite{Haldane83}. On a sphere, it is straightforward to verify that the energy levels are completely equivalent to the Haldane pseudopotential coefficients. On a torus, however, the pseudopotential algebra is not exact and so different eigenstates can mix, as well as potentially be effected by boundary conditions. In light of this, we examine the convergence of the energy levels in a two-particle system relative to the pseudopotential coefficients.

\begin{figure}
	\includegraphics[width=.7\linewidth]{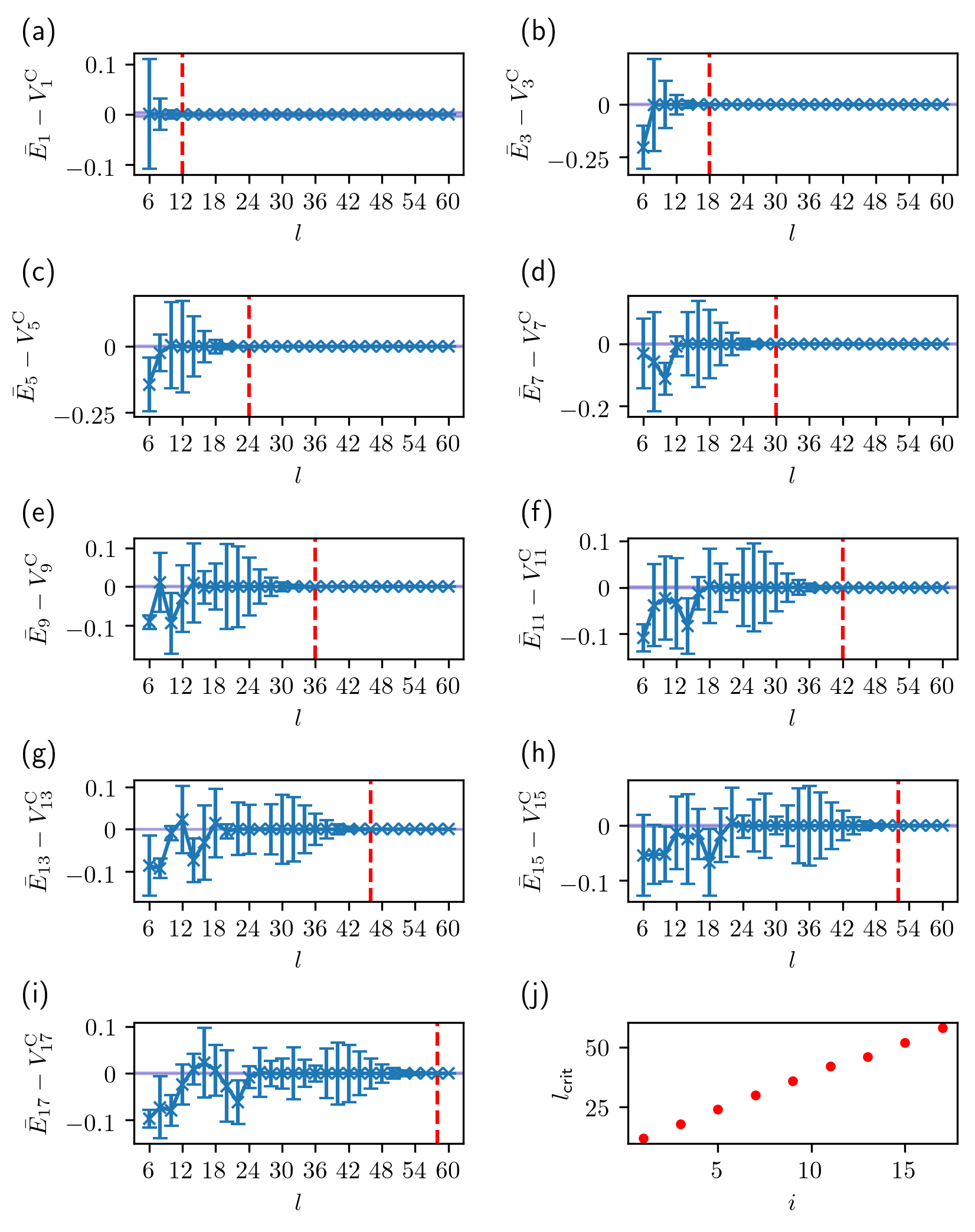}
	\caption{\label{fig:pp_torus}\textbf{Convergence of two-particle energy spectra to Haldane pseudopotential coefficients.} (a--i)~The set of quasi-degenerate energies $\{E_{i}\}$ corresponding to Coulomb pseudopotentials $V^\text{C}_i$, for the two-particle energy spectrum on a torus in the LLL, as a function of the number of flux quanta $l$. The mean $\bar{E}_i$ and standard deviation $\sigma_i$ of $\{E_{i}\}$ are plotted and the $V^\text{C}_i\pm0.01V^\text{C}_i$ region is shaded blue. The first value of $l$ for which $\bar{E_i}\pm\sigma_i$ is within $1\%$ of $V^\text{C}_i$ ($l_\text{crit}$) is marked with a red dashed line. (j)~Scaling of $l_\text{crit}$ as a function of Haldane pseudopotential coefficient $i$.}
\end{figure}

We consider a system on a torus with two particles in the LLL, with a size defined by the number of flux quanta $l$. Subsequently, we compute the energy spectrum of this system for a particular Haldane pseudopotential coefficient corresponding to the Coulomb interaction $V^\text{C}_i$. For each coefficient $V^\text{C}_i$, the resulting energy spectrum consists of a set of three quasi-degenerate energies $\{E_{i}\}$ with all other energies equal to zero. The plots of $\bar{E}_i-V^\text{C}_i$ as a function of $l$ are shown in Fig.~\ref{fig:pp_torus}. In particular, here we define convergence when the relative error $|\bar{E}_i-V^\text{C}_i|/V^\text{C}_i < 1\%$ and we denote the first value of $l$ where this occurs as $l_\text{crit}$.  

From Fig.~\ref{fig:pp_torus}, we can gain several insights into the effectiveness of the Haldane pseudopotential representation on a torus. First and foremost, we can see that the higher-order Haldane pseudopotential coefficients are more difficult to reproduce than for lower orders, with the required system size $l_\text{crit}$ scaling linearly with the order $i$. For example, although the first pseudopotential $V^\text{C}_1$ is already reproduced at $l_\text{crit}=12$, the third pseudopotential $V^\text{C}_5$ requires double this with $l_\text{crit}=24$, and so on. Second, we can conclude that for a given system with $N_\Phi$ flux quanta, the largest Haldane pseudopotential coefficient that should be employed in the interaction is $\beta \sim N_\Phi/2$. We note, however, that the precise value of this upper bound depends on system parameters, including the filling factor, and so cannot be precisely determined from Fig.~\ref{fig:pp_torus}. Physically, any coefficients larger than this threshold will be sensitive to the tails of the wave function wrapping around the cycles of the torus and so are severely effected by finite-size effects. Finally, we can see that at inadequate $l$ not all of the energies are resolved, and so the quasi-degenerate set of $\{E_{i}\}$ contains at least one zero, which pulls the average down unpredictably. For larger system sizes, we observe the expected symmetric and monotonic convergence of the mean and standard deviation towards the Haldane pseudopotential coefficient in all cases.

Building on this, we consider the case of Haldane pseudopotentials modeling the Yukawa interaction. From Fig.~\ref{fig:pp}, we can see that the form of the Yukawa pseudopotential drastically changes in the region $10^{-2}<\lambda<1$. At $\lambda=10^{-2}$, we observe a pseudopotential indistinguishable from the Coulomb pseudopotential on the scale of the plot, whereas at $\lambda=1$ we observe a sharp cut-off, where only the first few pseudopotentials take significant non-zero values. Due to this sharp transition, it is no longer critical to consider higher-order pseudopotentials in order to accurately model the Yukawa interaction at large $\lambda$. Since the analysis in Fig.~\ref{fig:pp_torus} is performed one pseudopotential at a time, the results are simply scaled for the Yukawa interaction. Hence, the relative error thresholds are unaffected and the scaling in Fig.~\ref{fig:pp_torus}(j) still holds. However, since the higher-order pseudopotentials are significantly smaller for the Yukawa interaction with large $\lambda$, the absolute error threshold $|\bar{E}_i-V_i| < 1\%$ is satisfied at drastically smaller system sizes.

In summary, although the Haldane pseudopotentials have a complete equivalence with their two-particle energy spectra on a sphere, this does not hold for the torus. On a torus geometry with $N_\Phi$ flux quanta, the Haldane pseudopotentials can only be used to model short-range interactions described by $\{V_1, V_3, \dots, V_{\beta}\}$, where $\beta \ll N_\Phi$.

\subsection{Tuning from lower- to higher-order Haldane pseudopotentials}
\label{sec:trunc}


To complement our analysis of the Coulomb and Yukawa interactions using an explicit diagonalization in Fourier space, we study the interactions using a Haldane pseudopotential description. The Haldane pseudopotential formalism is useful because most of the physics of interacting fermions in the LLL can be captured by its first few values, which greatly simplifies both analytical and computational complexity. Although originally applied to short-range interactions on the sphere, it has since been generalized to accommodate more diverse interactions on a range of geometries~\cite{Pakrouski17}. Nevertheless, it is important to note that, unlike for the sphere, pseudopotential algebra is not exact on the torus and so a convergence of the energy spectrum will only be reached for the first few energy levels with short-range interactions relative to the system size.     

\begin{figure}
	\includegraphics[width=.7\linewidth]{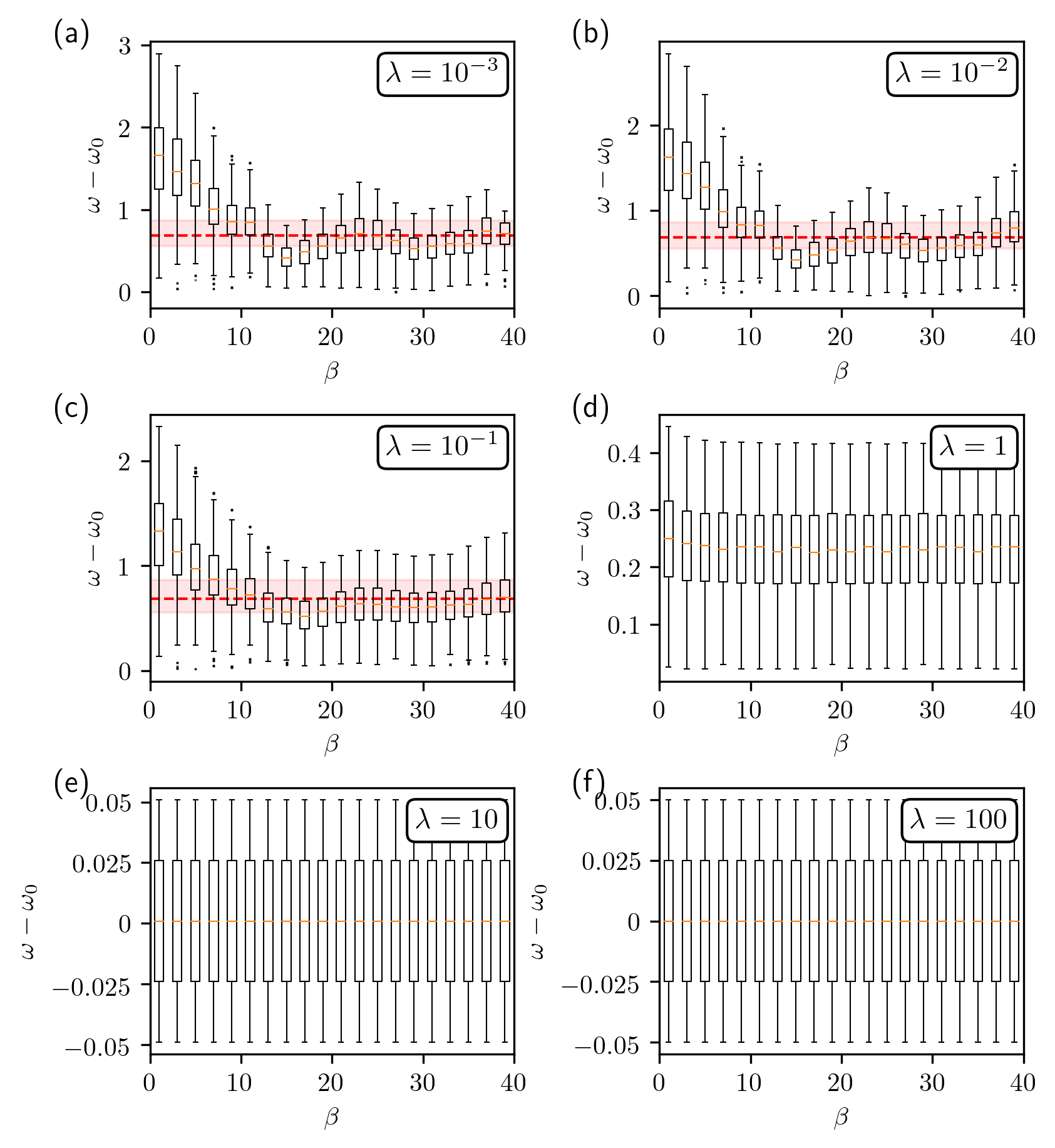}
	\caption{\label{fig:ypt}\textbf{Convergence of the structure factor with Yukawa pseudopotential truncation.} Box plots showing the spread of the structure factor $I_{\bar{\rho}}(0, \omega)$ for the $\nu=1/3$ FQH state on a torus, with $N_f=6$ particles and $N_\Phi=18$ flux quanta, stabilized by the LLL Haldane pseudopotentials corresponding to the Yukawa interaction $\{V^\text{Y}_{\lambda}\}=\{V^\text{Y}_{\lambda,0}, V^\text{Y}_{\lambda,1}, \dots, V^\text{Y}_{\lambda,\beta}\}$, as a function of truncation parameter $\beta\in\{1,3, \dots, 39\}$ at a variety of $\lambda$. The median is labeled with an orange line, the interquartile range (IQR) is drawn with a box, and the whiskers extend to 1.5 times the IQR. All data points outside of this range are plotted as outliers. In (a--c) we additionally overlay the IQR and median for the exact Coulomb interaction from Fig.~1(b) in red. The computations were performed with a resolution of $\Delta\omega=10^{-5}$, $\Delta I =10^{-5}$, and $\epsilon=10^{-4}$.}
\end{figure}

In order to quantify the effectiveness of the first few Haldane pseudopotentials in capturing the physics of the Laughlin state on the torus, we present the scaling of the structure factor as we increase the order of the pseudopotentials corresponding to the Yukawa interaction $\{V^\text{Y}_{\lambda}\}=\{V^\text{Y}_{\lambda,0}, V^\text{Y}_{\lambda,1}, \dots, V^\text{Y}_{\lambda,\beta}\}$ in Fig.~\ref{fig:ypt}, where $\beta$ is our truncation parameter. Although, it is physical to consider only small $\beta\ll N_\Phi$, since larger values may reflect eigenstate mixing and boundary conditions, we show an extended range of $\beta\in\{1,3, \dots, 39\}$ to additionally comment on numerical effects. In Fig.~\ref{fig:ypt}(a), we show the structure factor scaling in the Coulomb approximation $\lambda\approx0$. From the plot, we can see that for $\beta\gtrsim N_{\Phi}/2$, we obtain a rough approximation of the structure factor corresponding to the Coulomb interaction. However, we note that for such a large number of pseudopotentials, the convergence of the energy levels is limited, which results in oscillations of the energy scale. These oscillations reflect the boundary interference arising from modeling large relative angular momenta on a finite torus. As we increase the value of $\lambda$, and hence decrease the interaction range, these boundary effects are gradually alleviated. Comparing Fig.~\ref{fig:ypt}(c) with Fig.~\ref{fig:ypt}(a), for example, we can see that the oscillations for $\beta\gtrsim N_{\Phi}/2$ are diminished and the approximation to the Coulomb interaction at $\beta\sim N_{\Phi}/2$ is improved. Further increasing the value of the Yukawa mass beyond $\lambda\sim1$ exponentially decreases the range of the interaction and consequently, suppresses the domain of the structure factors. In the limit of $\lambda\to\infty$, the mean and standard deviation tend to zero, as we saw in Fig.~1(e). These results stress that Haldane pseudopotentials can work well on the torus \emph{provided} a few compromises are reached. The number of pseudopotentials needs to be large enough to accurately model the interaction potential, but not so large as to introduce boundary effects and eigenstate mixing. We find that, for this system, the optimal number of pseudopotentials is $\beta\sim N_{\Phi}/2$. Moreover, long-range interactions, such as the Coulomb interaction are more susceptible to finite-size effects. Therefore, the structure factor for these systems is most accurately modeled using Haldane pseudopotentials corresponding to weakly screened forms of the interaction.

\begin{figure}
	\includegraphics[width=.7\linewidth]{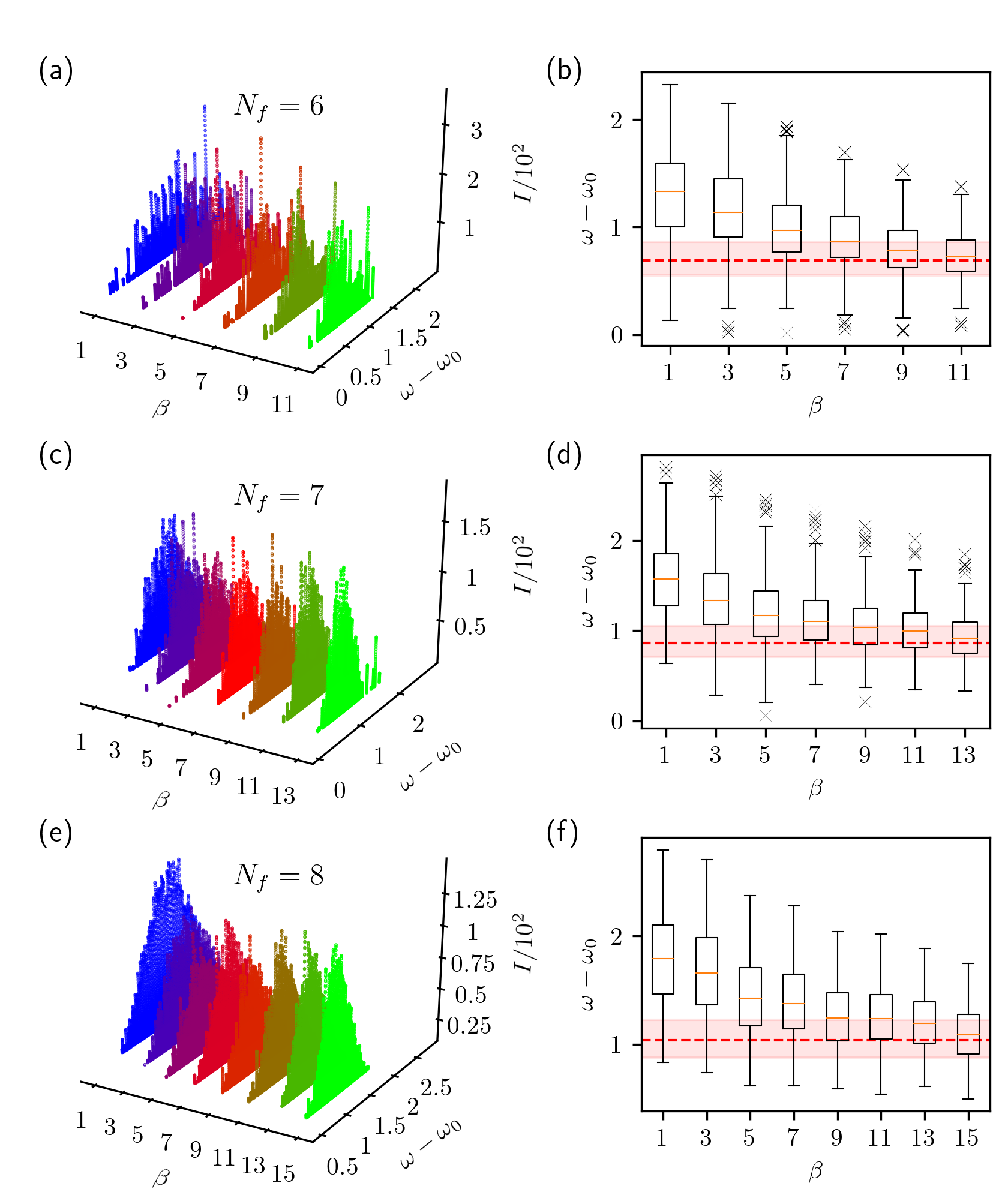}
	\caption{\label{fig:ypt_l1}\textbf{Approximating the Coulomb structure factor using Yukawa pseudopotentials at \mbox{$\lambda=0.1$}.} Structure factor $I_{\bar{\rho}}(0, \omega)$ for the $\nu=1/3$ FQH state on a torus, corresponding to the system in Fig.~\ref{fig:ypt}(c). (a,c,e)~3D plots showing the scaling of $I_{\bar{\rho}}(0, \omega)$ with truncation parameter $\beta\in\{1,3,\dots, \beta_\mathrm{max}\}$, where $\beta_\mathrm{max}$ is the smallest odd integer greater than $(N_{\Phi}+2)/2$, for (a)~$N_f=6$, (c)~$N_f=7$, and (e)~$N_f=8$. (b,d,f)~Box plots showing the spread of $I_{\bar{\rho}}(0, \omega)$ as a function of $\beta$, for (b)~$N_f=6$, (d)~$N_f=7$, and (f)~$N_f=8$. Outliers are shown explicitly as black crosses. The center and spread for the corresponding exact Coulomb distributions are overlaid in red. The computations were performed with a resolution of $\Delta\omega=10^{-5}$, $\Delta I =10^{-5}$, and $\epsilon=10^{-4}$.}
\end{figure}

Having shown that Haldane pseudopotentials can successfully model the Coulomb interaction provided the interaction is sufficiently screened, we now focus on the approximation obtained using the Yukawa interaction with $\lambda=0.1$. In Fig.~\ref{fig:ypt_l1}, we present a finite-size scaling of the structure factor from Fig.~\ref{fig:ypt}(c) for the particle numbers $N_f=6,7,8$ and the physical values of the truncation parameter $\beta\ll N_{\Phi}$. From Fig.~\ref{fig:ypt_l1}(a), we find a distribution with a single distinct mode akin to Fig.~1(b), with a comparable number of peaks and spread. Here, we observe the optimal approximation to the structure factor for the Coulomb interaction at $\beta=11$, as shown in Fig.~\ref{fig:ypt_l1}(b). As we increase the particle number in Figs.~\ref{fig:ypt_l1}(c--f), the number of pseudopotentials required for an accurate approximation also increases, such that optimal number is consistently $\sim N_{\Phi}/2$. This is the threshold that maximizes the accuracy of the interaction potential before boundary effects from the finite torus manifest, however we note that the precise value is dependent on system parameters, such as the filling factor. In addition to this, there are a few notable trends as we increase the system size. First, as was suggested by the results in Fig.~2, we find that the peak fluctuations are suppressed with increasing system size, where we no longer observe any outliers in the box plots in Fig.~\ref{fig:ypt_l1}(f), compared to Figs.~\ref{fig:ypt_l1}(b,d), for example. Second, we observe that for a restricted set of pseudopotentials, the approximation is more accurate for smaller system sizes. For example, the $\beta=1$ approximation for $N_f=6$ is more accurate, in relative terms, than the $\beta=1$ approximation for $N_f=8$. Finally, we can see that by comparing Fig.~\ref{fig:ypt_l1}(a) and Fig.~1(b), for instance, the features near the bulk of the response spectra are reproduced more rapidly than the structure at the edges. These points notwithstanding, the structure factors presented here are a fair approximation of that for the long-range Coulomb potential in Fig.~1(b) and come at a significantly reduced numerical cost.

\section{Details of the Lanczos algorithm}
\label{sec:lanczos}

The underlying motivation behind the Lanczos algorithm is to compute the eigendecomposition of a Hermitian Hamiltonian $H$ by variationally minimizing the energy functional $E[\Psi]=\braket{\Psi|H|\Psi}/\braket{\Psi|\Psi}$. Progress is made towards this goal by using an iterative procedure that subsequently explores the relevant parts of the Hilbert space.

In its most basic form, the Lanczos algorithm takes a normalized input vector $\ket{v_0}=\ket{\Psi}/\sqrt{\braket{\Psi|\Psi}}$, orthonormalizes $H\ket{v_0}$ with respect to $\ket{v_0}$ to obtain $\ket{v_1}$, and then finds the variational state of lowest energy by diagonalizing $H$ in the two-dimensional subspace $\text{span}(\ket{v_0}, \ket{v_1})$. This variational state of lowest energy is then used as the input for the next minimization step, and so on, until the desired eigenenergy converges.

Building on this idea, efficient implementations directly perform $M-1$ such iterations by finding the variational state of lowest energy in the $M$-dimensional Krylov space $\mathcal{K}^{M-1}(\ket{v_0})=\text{span}(\ket{v_0}, H\ket{v_0}, H^2\ket{v_0}, \dots, H^{M-1}\ket{v_0})$, where $\{\ket{v}\}$ is the set of Lanczos vectors. As before, the algorithm starts with a normalized input vector $\ket{v_0}$. It then orthogonalizes $H\ket{v_0}$ with respect to $\ket{v_0}$ and normalizes to yield $\ket{v_1}=\ket{\tilde{v}_1}/\sqrt{\braket{\tilde{v}_1|\tilde{v}_1}}$, where $\ket{\tilde{v}_1} = (1-\ket{v_0}\bra{v_0})H\ket{v_0}$. Similarly, $\ket{\tilde{v}_2}$ is constructed by orthogonalizing $H\ket{v_1}$ with respect to the two previous Lanczos vectors, such that $\ket{\tilde{v}_2}=H\ket{v_1} - a_1 \ket{v_1} - b_1 \ket{v_0}$, where $a_n=\braket{v_n|H|v_n}$ and $b_n=\bra{v_{n-1}} H \ket {v_n}$. From the second step onwards, the iteration proceeds in the same manner, by orthonormalizing $H\ket{v_n}$ with respect to $\ket{v_n}$ and $\ket{v_{n-1}}$, i.e.~using  $\ket{\tilde v_{n+1}} = H\ket{v_n} - a_n \ket{v_n} - b_n \ket{v_{n-1}}$ followed by $\ket{v_n}=\ket{\tilde{v}_n}/\sqrt{\braket{\tilde{v}_n|\tilde{v}_n}}$. Crucially, due to the Hermiticity of the Hamiltonian $H$, as well as the orthogonality of the Lanczos vectors $\{\ket{v}\}$, the resulting basis is orthogonal (to within numerical precision) and only the current and immediately preceeding Lanczos vectors are needed to compute the subsequent vector. This means that the Hamiltonian may be expressed as
\begin{equation}
H\ket{v_n} = b_n \ket{v_{n-1}} + a_n \ket{v_n} + b_{n+1} \ket{v_{n+1}}.
\end{equation}
Therefore, in matrix form, the original $N\times N$ Hermitian Hamiltonian matrix $H$ can be represented approximately by a $M\times M$ tridiagonal real symmetric Lanczos Hamiltonian matrix $\check{H}$ in Krylov space, given as
\begin{equation}
\check{H}=
\begin{pmatrix}
a_0 & b_1 &  &  & \\
b_1 & a_1 & b_2 &  &  \\
& b_2 & a_2 &\ddots &  \\
&  &\ddots & \ddots & b_{M-1} \\
&  &  & b_{M-1} & a_{M-1}
\end{pmatrix},
\end{equation}
where $M\leq N$.

Although the Lanczos algorithm does not compute the eigendecomposition of $H$, it takes a significant step towards this goal, by yielding a tridiagonal matrix $\check{H}$ that can be readily diagonalized. Moreover, since the Lanczos algorithm is an iterative method, useful results can be obtained without the algorithm running to completion. Most pertinently, extremal eigenvalues of $\check{H}$ show good agreement with those of the original system, even when the number of Lanczos iterations is much smaller than the dimension of the original Hamiltonian, $M \ll N$~\cite{Bai00, Koch11, Dargel12}.

\section{Derivation of the momentum-averaged density operator on a torus}
\label{sec:form_fac}

Consider a free charged fermion in the $xy$-plane in the presence of a perpendicular magnetic field $\mathbf{B}=B\hat{\mathbf{z}}$. The fermion motion is typically parameterized using its center-of-mass coordinates $(X, Y)$ and characterized by the magnetic length $l_B$. In Landau gauge with a conserved $y$-momentum, $k_y$, it can be shown that the single-particle ground states may be written as
\begin{equation}
	\label{eq:single}
	\phi(x,y)\sim e^{\mathrm{i}k_y y} \exp\left( -(X-x)^2/2l_B^2 \right),
\end{equation}   
where $X=-k_yl_B^2$. This shows that the wavefunctions are localized in the $x$-direction but extended in $y$.

When the fermion is confined to a rectangular sample of sides $L_x$ by $L_y$, then the degeneracy of the LLL is determined by the number of allowed $k_y$ such that $0\leq X<L_x$. Applying periodic boundary conditions to Eq.~\eqref{eq:single}, we obtain $k_y = X_j/l_B^2 = 2\pi j / L_y $, where $j = 0,\ldots,N_\Phi-1$ is an integer bounded by the degeneracy of the LLL, $N_\Phi = L_x L_y / 2\pi l_B^2$. There are consequently $N_\Phi$ LLL eigenstates, which may be written as
\begin{equation}
	\label{eq:LandauOrbsTorus}
	\phi_{j}(x, y)\sim \sum_{m=-\infty}^\infty e^{\mathrm{i}(X_j + m L_x)y/l_B^2} \exp\left( -(X_j + mL_x -x)^2/2l_B^2 \right),
\end{equation}      
where $0\leq j < N_\Phi$. The degeneracy is equivalent to the total number of flux quanta, such that $N_\Phi\equiv \Phi/\Phi_0$, where $\Phi$ is the magnetic flux and $\Phi_0$ is the flux quantum. From now on, we set the magnetic length to one $l_B=1$.

In order to compute the density operator on the torus we expand in the basis of LLL orbitals, such that
\begin{equation}
\label{eq:rho_momentum_expansion}
	\rho_{\mathbf{q}} = \sum_{j, j'} \hat{\rho}_{j,j'}(\mathbf{q}) c^\dagger_j c_{j'}, 
\end{equation}
where $c^{(\dagger)}_j$ are the annihilation (creation) operators for the Landau level orbitals $\phi_j$~\eqref{eq:LandauOrbsTorus} and $\hat{\rho}_{j,j'}(\mathbf{q})$ is the Fourier transform of the normalized particle density coefficients, or form factor, defined as
\begin{equation}
	\hat{\rho}_{j,j'}(\mathbf{q}) \equiv \int \mathrm{d}\mathbf{r} \, e^{\mathrm{i} \mathbf{q}\cdot\mathbf{r}} \hat{\phi}_j(\mathbf{r}) \hat{\phi}^*_{j'}(\mathbf{r}),
\end{equation}
$\mathbf{r}\equiv(x,y)$ and $\mathbf{q}\equiv(q_x,q_y)$ are position and momentum conjugate variables, and the hats denote normalization.

Starting with the form factor, we compute the Fourier transform of
\begin{equation}
	\label{eq:density}
	\hat{\rho}_{j,j'}(x,y) = \frac{\phi_{j}(x, y) \phi^*_{j'}(x,y)}{N_f},
\end{equation} 
where $N_f$ is a normalization constant, corresponding to the total particle number.

Computing the denominator of Eq.~\eqref{eq:density}, the total particle number
\begin{equation}
	N_f = \sum_j \int_0^{L_x} \mathrm{d} x \int_0^{L_y} \mathrm{d}y \,\phi^*_{j}(x, y) \phi_{j}(x,y),
\end{equation}
yields, in terms of the number of repetitions of the simulation cell in the $x$- and $y$-directions, $N_x, N_y$,
\begin{equation}
	\label{eq:N}
	N_f = \frac{1}{N_x N_y} \sum_j \int \mathrm{d}x \mathrm{d}y \,\phi^*_{j}(x, y) \phi_{j}(x,y).
\end{equation}
Working with large $N_x, N_y$ allows us to extend the integration range. Performing the $y$-integral yields
\begin{equation}
	N_f = \frac{2\pi L_y}{N_x L_x} \sum_{m=-\infty}^\infty \int \mathrm{d}x \exp\left( -\frac{1}{2}x^2-\frac{1}{2}(x-m L_x)^2\right),
\end{equation}
where we have invoked the momentum periodicity in the $y$-direction. Subsequently, performing the $x$-integral yields
\begin{equation}
	N_f = 2\pi\sqrt{\pi}L_y \sum_{m=-\infty}^\infty \exp\left( -m^2 L_x^2/4\right).
\end{equation}

In a similar fashion, we can compute the Fourier transform of the numerator of Eq.~\eqref{eq:density}, the unnormalized particle density coefficients
\begin{equation}
	\rho_{j,j'}(\mathbf{q})=\frac{1}{N_x N_y}\int \mathrm{d}\mathbf{r}\, e^{i \mathbf{q}\cdot\mathbf{r}} \phi_{j}(\mathbf{r}) \phi^*_{j'}(\mathbf{r}). 
\end{equation}
As before, working with large $N_x$, $N_y$ allows us to extend the integration range, and we can ultimately take $N_i\to\infty$ to obtain complete Gaussian integrals along $x$. Performing the $y$-integral yields
\begin{equation}
	\rho_{j,j'}(\mathbf{q}) = \frac{2\pi L_y}{N_x L_x}\delta(X_j - X_j' +q_y) \sum_{m=-\infty}^\infty \int \mathrm{d}x\, e^{\mathrm{i} q_x x}\exp\left( -\frac{1}{2}(x-X_j')^2 -\frac{1}{2}(x-X_j-m L_x)^2\right),
\end{equation}
where we have again employed the momentum periodicity in the $y$-direction. Finally, evaluating the integral along $x$ yields
\begin{equation}
	\rho_{j,j'}(\mathbf{q}) = 2\pi \sqrt{\pi} L_y \delta(X_j - X_j' +q_y) \sum_{m=-\infty}^\infty \exp[ \chi_{j,j',m}(q_x)],
\end{equation}
where
\begin{equation}
	\chi_{j,j',m}(q_x) \equiv \left(\frac{X_j'+X_j+mL_x+\mathrm{i}q_x}{2} \right)^2 - \frac{(X_j+mL_x)^2}{2}-\frac{X_j'^2}{2},
\end{equation}
and hence the form factor is given as
\begin{equation}
	\hat{\rho}_{j,j'}(\mathbf{q}) =
	\delta(X_j - X_j' +q_y) \frac{\sum_{m=-\infty}^\infty \exp \chi_{j,j',m}(q_x)}{\sum_{m=-\infty}^\infty \exp\left( -m^2 L_x^2/4\right)}.
\end{equation}
By invoking the $q_x$ periodicity of the form factor, we can integrate out the $q_x$ dependence by summing over momentum modes on the torus. This leaves us with an expression for the density operator that is only dependent on the $q_y$ modes, which has analogous properties to the conserved $k_y$ momentum in Landau gauge. Hence, using these coefficients in the general expansion of the density operator~(\ref{eq:rho_momentum_expansion}), we arrive at our final expression for the $q_x$-momentum-averaged density operator used in the main text
\begin{equation}
	\bar{\rho}_{q_y} = \sum_{m=0}^{N_\Phi - 1} \rho_{q_x = \frac{2\pi m}{L_x}, q_y}.
\end{equation}

\bibliographystyle{apsrev4-1}
\bibliography{ms}